\documentclass[twocolumn]{article}
\usepackage{graphicx}
\usepackage{amsmath}
\usepackage{caption}
\usepackage{authblk}
\usepackage{geometry}
\usepackage[english]{babel}
\title{Increasing  the link-distance of  free-space quantum coherent communication with large area detectors
}

\author[1,2]{Rupesh Kumar}
\author[2]{Igor Konieczniak }
\author[3]{Gerald Bonner }
\author[1,2]{Tim Spiller}

\affil[1]{Quantum Communications Hub, University of York, Heslington, YO10 5DD, UK}
\affil[2]{ Department of Physics, University of York,Heslington, YO10 5DD, UK}
\affil[3]{Fraunhofer Centre for Applied Photonics, Fraunhofer UK Research Ltd, 99 George Street, Glasgow, G1 1RD, UK

}
\date{}

\begin{document}

\twocolumn[
  \begin{@twocolumnfalse}
    \maketitle
    
\vspace{-6mm}
\begin{abstract}
We report a large area photo-diode based homodyne detector for free-space quantum coherent communication. The detector's performance is studied in terms of detection bandwidth and electronic noise for shot-noise limited quantum signal detection. Using large area photo-diodes increases signal collection efficiency from turbulent atmospheric channels, in comparison with  typical fibre based free-space homodyne detectors. Under identical atmospheric turbulence and receiver  aperture conditions, our homodyne detector based on 1mm diameter photo-diode experiences 0dB loss due to turbulence while a 10um fibre based detector  experiences  13.5dB of signal loss over a 700km free-space link, at 90 degree elevation angle.
\end{abstract}
  \end{@twocolumnfalse}
]

\section{Introduction}
In coherent communication systems,  optical signals carry information on both their amplitude and phase\cite{Kikuchi2016}.  Coherent detection measures the amplitude and phase  of the  signal in terms of the quadrature  with respect to a strong reference signal called the local oscillator (LO).  Since coherent detectors  can differentiate multiple values of signal quadratures, coherent communication has a higher data carrying capacity compare to threshold detection approaches, for example on-off-keying (OOK), where signal state discrimination is limited to two or a few values.   At the lowest intensity levels, near to vacuum, coherent signals evidently show inherent quantum uncertainty in their quadratures. This prevents the different quadrature values being effectively discriminated from each other and hinders data communications. Nevertheless coherent signals at the quantum level are useful for secure key exchange  between two authenticated users\cite{Gisin2002,Diamanti2015,xu2020}.   However the coherent detector needs to have vacuum noise (shot-noise) sensitivity to differentiate signals at the quantum level. Quantum coherent communication for secure key exchange, referred to as continuous variable quantum key distribution (CV-QKD)\cite{Grosshans2002}, exploits the vacuum noise of the quanutm signals to ensure security of the key.  In CV-QKD the quadratures values sent by the transmitter and measured by the receiver are correlated, if properly measured. This correlation can be established through subsequent data reconciliation. Thereafter, error correction followed by privacy amplification are employed to generate a secure key\cite{Jouguet2011}.

Degradation in correlation between transmitted and measured quadratures indicates the presence of noise and loss. Vacuum noise  of the signal and electronic noise of the detector are two sources of noise that can be calibrated in CV-QKD, whilst the noise from eavesdropping limits the security of the key generation. Attenuation due to the channel and in the receiver, together with detector inefficiency, contribute to the signal loss. In fibre-based systems, the signal is confined in the core of the fibre and signal attenuation is primarily due to absorption/scattering. Many techniques have been proposed and are being developed in order to overcome the losses and to extend the transmission distance, such as measurement device independent  (MDI)-QKD\cite{Braunstein2012, Lo2012}, use of quantum repeaters\cite{Duan2001}  and, most recently, Twin Field (TF)-QKD \cite{LucamariniTF2018}. All of these methods require an intermediate node(or nodes) between the users. 

However, for free-space QKD channels it is quite impractical to extend the overall working link distance by placing a node in between, especially for satellite to ground links\cite{Bedington2017}. In this paper we are proposing an efficient way to improve the signal collection efficiency in free-space CV-QKD systems, especially for transmitted local oscillator(TLO) based systems. We demonstrates a free-space homodyne detection system with large area photo-diodes and  evaluate its effect in CV-QKD operational bandwidth and electronic noise. We show that using larger area detectors,  it is possible to achieve higher signal collection efficiency for CV-QKD in atmospheric channels.  
We have the following sections in this paper. In section 2, we describe the shot-noise sensitive homodyne detection  for coherent signals. Section 3 describes the relationship of detector area with bandwidth and electronic noise.  In Section 4 we examine the advantage of a large area detector in signal detection compared to single mode fibre based detectors and provide  evidence in terms of shot-noise variance and linear detection response as a preliminary result for conducting free-space CV-QKD demonstration. We conclude the paper in Section 5.

\section{CV-QKD over free-space atmospheric channel}

 Free-space CV-QKD has recently gained interest  primarily because of its capability to filter out noise photons which are not coherent with the LO. This enables daylight operation of QKD  in satellite-ground links\cite{Gunthner2017,Guo2018, Hosseinidehaj2019}. A few theoretical studies have been carried out in the direction of increasing the loss budget of CV-QKD systems in atmospheric channels. In \cite{Ghalaii2019}, a realistic and passive eavesdropping scenario is explored, whilst in \cite{Gunthner2017} a virtual aperture is considered as the transmitter of the quantum signals. In this paper, we propose a novel concept in quantum coherent signal detection, in order to improve the signal collection efficiency for CV-QKD in atmospheric channels. We also provide preliminary test results to support this proposal.

In a free-space channel, signal loss is not only caused by atmospheric absorption and scattering but also from the beam divergence and by the atmospheric turbulence. Beam divergence is the
geometrical spreading of the signal due to diffraction. For short range free-space communications, say a few tens of km, it is possible to keep the beam width of the signal smaller than the receiver aperture. As a result, loss due to beam divergence is negligible. However at longer transmission distances, as in the case of  satellite to ground links, since the beam width  becomes much larger than any practical receiver aperture, only a portion of the signal can in fact be collected.

The portion of the signal collected by the receiver telescope is further coupled to the detector, either via free-space or with fibre pigtails. It is a demanding task as it requires precise pointing, acquisition and tracking (PAT) subsystems and signal focusing optics in order to keep the signal focused onto the detector area - which is typically 10µm in single mode fibre, 80µm in multi-mode fibre and 200µm in free-space detector.  Random variations of the refractive index of the atmosphere generate scintillation effects that results in signal fading, wavefront distortion and random motion of the signal beam centroid about the receiver. After all, the atmospheric turbulence spreads the signal spot size  and reduces the signal collection efficiency significantly. Adaptive optical techniques can correct the wavefront distortion and increase the signal coupling efficiency to the detector with an aid from beacon signal, at a separate wavelength, that is transmitted along with the QKD signal\cite{Gunthner2017}. Precision telescopic mounts with high tracking resolution can follow the wandering signal beam. However, this compromises in tracking speed\cite{Baister1994}.

\section{Quantum coherent signal detection}

Since the signal strength is comparatively higher, a classical coherent detector has diverse detection capabilities - in terms of orthogonal polarizations for both quadratures. On the contrary, quantum coherent signals are very weak in intensity - a few photons per pulse on average at the transmitter\cite{Jouguet2011}. Diversified signal detection over a lossy channel creates further loss. Therefore, quantum coherent signals are prepared in a single polarization and detected preferably in a single quadrature. Single quadrature detection is referred to as homodyne detection, whilst detection of both quadratures is termed heterodyne. Based on the protocol for secure key generation, the CV-QKD detector either randomly measures one of the quadratures, or measures both quadratures. 
We consider  a homodyne detector  in our analysis primarily because of its use in a well studied and demonstrated Gaussian modulated coherent state (GMCS) protocol\cite{Grosshans2002, Jouguet2011} for CV-QKD. 

\begin{figure}[htb!]
 \includegraphics[width=0.5\textwidth]{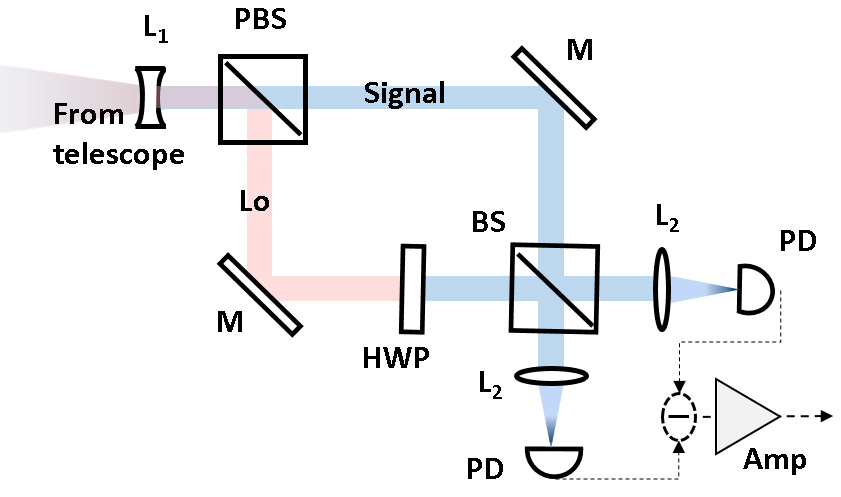}
\caption{Set-up for free-space homodyne detection. Focused signal from the primary mirror of the telescope is first collimated  by  lens L1 and then passes through the polarization beam-splitter (PBS) for separating orthogonal polarized signal and lo. Polarization of the LO is rotated with half wave plate (HWP) to match with that of signal. The signal and LO interfere at a 50/50 beam-splitter (BS) and its outputs are each focused to a photo-diode (PD) using lens L1 and L2. M stands for the mirror.  }
\label{HD}     
\end{figure}

In  homodyne detection for quantum signals, the signal field $ \alpha e^{i\theta_s}$ is mixed with a strong LO field $\sqrt{I_{lo}}e^{i\theta_{lo}}$ on a symmetric beam-splitter  as shown in figure \ref{HD}. Here $\alpha$  is the complex amplitude  and  $I_{lo}$ is the intensity of the  LO.   The signal can be expressed in terms of field quadratures, $ X+iP$. Since, the LO and signal originate from the same laser source in TLO based CV-QKD systems,  the signal quadrature with respect to the LO can be written as $ (X_{\theta_{lo}} + X_0 )+i (P_{\theta_{lo}}+P_0)$. 
Here, $X_0$ and $P_0$ are the random quadrature values correspond to vacuum noise variance - which does not have any phase relation with the LO.   The outputs of the beam-splitter are connected to  photo-diodes.  The photo-currents are subtracted from each other  and the difference is amplified to a detectable level. The amplifier  output directly indicates the quadrature of the input signal where the relative phase of the LO with respect to the signal determines the quadrature under measurement.

The homodyne detection output, $HD_{out}$, for an $X$ quadrature measurement (and similarly for the $P$ quadrature) can be written as:
\begin{equation}
HD_{out}= 2\sqrt{I_{lo}}(X_{\theta_{lo}} + X_0 ) + X_{ele}
\label{HD}
\end{equation}

where, $X_{ele}$ is the noise quadrature contribution due to the electronic noise variance  $V_{ele}$. The signal quadrature $X_{\theta_{lo}}$ may contain excess noise that originates from other sources, which we are neglecting in our analysis.  The shot-noise limited sensitivity of the homodyne detector is obtained by  using a high intensity LO and low electronic noise generating components - photo-diodes and amplifier. Primarily, the electronic noise variance determines the intensity required for the LO to bring the homodyne detector to shot-noise sensitivity. Therefore, it is important to evaluate how electronic noise varies with different aspects of the detector.

\section{Homodyne detector with large area photo-diodes}
 
 In general, there are three figures of merit for a receiver:  detection efficiency, bandwidth  and electronic noise.  Detection efficiency  is the product of the quantum efficiency  of the  photo-diodes  and the optical loss in  the receiver including the telescope. The detection bandwidth of the receiver  is bounded by the electrical bandwidth of the photo-diode and the  amplifier.
This is relevant as it can limit the signal repetition rate. Typically  repetition rate is set to 1/3 of the bandwidth and signal pulse width is to 10$\%$ of the duty-cycle\cite{Xinke2018}. In order to make the homodyne detector shot-noise sensitive, it is required to have comparatively low electronic noise from the detection electronics - $3^{rd}$ term in Eq.\eqref{HD}. Its contribution can be made smaller either  by  higher LO intensity, $I_{lo}$ - $1^{st}$ term in Eq.\eqref{HD}, or  using low noise electronic components at the expense of reduction in the bandwidth. In a homodyne detector, the bandwidth, $B$, and electronic noise variance, $V_{ele}$,  are related by the following equation \cite{Xinke2020}. 
 \begin{equation}
V_{ele} = \frac{NEP^2 B \tau }{ h \nu I_{lo}}
\label{BW}
\end{equation}
where,  $NEP$ is the  sum of the noise-equivalent  power  of the  photo-diodes and the amplifier, $\tau$ is the LO pulse width, $h$ is Planck's constant and $\nu$ is the optical LO frequency.  
\begin{figure}[htb!]
 \includegraphics[width=0.5\textwidth]{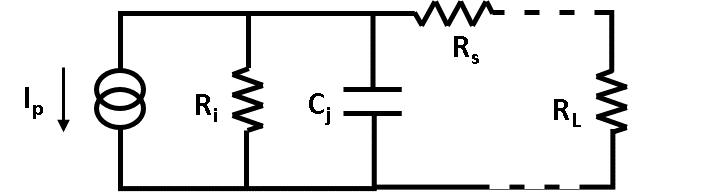}
\caption{ Equivalent circuit of a photo-diode. $C_j$ is the junction capacitance, $R_i$ is the internal resistance and $R_s$ is the series resistance, $R_L$ is the load resistance and $I_p$ is the photocurrent.}
\label{PD}     
\end{figure}
In our noise model, we consider the NEP of the amplifier to be constant and analyse how the detection area of the photo-diode affects the NEP, bandwidth and thence the electronic noise variance. In order to relate the area of the  photo-diode to bandwidth, we consider the equivalent circuit of a photo-diode as shown in figure \ref{PD}.  For a photo-diode with junction capacitance $C_j$ connected to a load resistor, $R_L$, the bandwidth $B=1/2 \pi R_L C_j $ can be expanded as: 

\begin{equation}
B = \frac{1}{\pi R_L A}\sqrt{\frac{\mu \rho (V_A + V_{bi})}{2 \epsilon \epsilon_0}}
\label{BW_Area}
\end{equation}
where $A$ is the detection area of the diode, $\mu$ is the mobility of electrons at 300K, $\rho$ is the resistivity of the photo-diode material, $V_A$ is the built in potential of the material, $V_{bi}$ is the applied reverse bias voltage, $\epsilon$ is the dielectric constant and $\epsilon_0$ is the permittivity of free space. As the area of the detector increases the detection bandwidth reduces, due to increase in junction capacitance, as shown in figure \ref{BW}.
\begin{figure}[htb!]
 \includegraphics[width=0.5\textwidth]{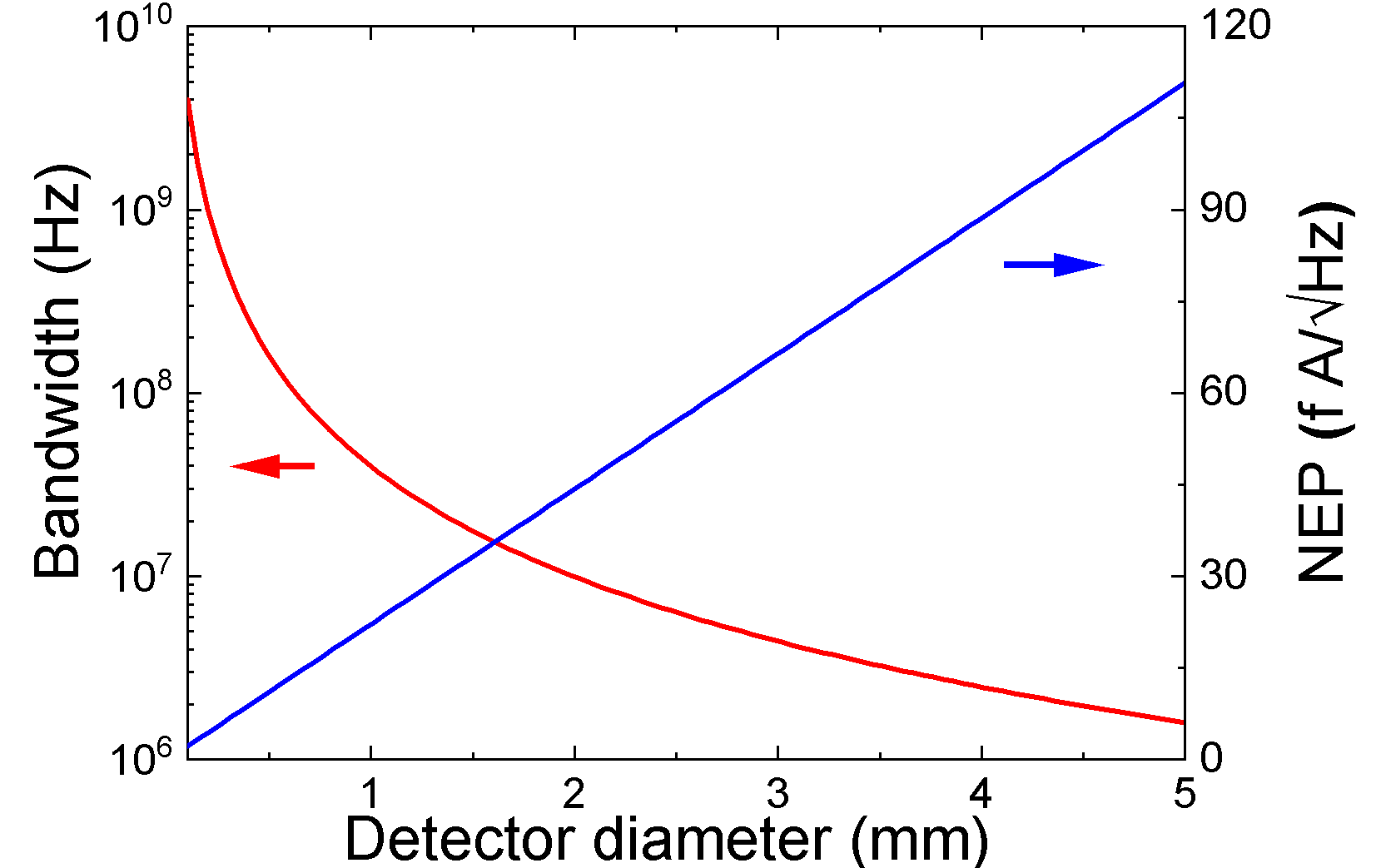}
\caption{ Variation of the bandwidth (red line)  and $NEP$ (blue line)  with respect to the diameter of the photo-diode. 
The following values are used in Eq.\eqref{BW_Area} and Eq.\eqref{NEP}: $R_L=50\Omega$; $\mu=10^4\frac{cm^2}{V s}$; $\rho=0.142cm\Omega$; $V_A=0.77V$; $V_{bi}=6V$; $\epsilon=13.9$; $n_i=6.3\times10^{11}cm^{-3}$; $N_A=1.2\times 10^{31}cm^{-3}$;$T=300K$; $\mu_n= 250 cm^2/(Vs)$; $\tau_n=270\times 10^{-15}s$; $t=55cm$; $L_n=14nm$; $\gamma=95\%$.}
\label{BW}     
\end{figure} 

 Considering the dark current as the source of noise in the photo-diode, the $NEP$ can be defined as the power required to generate a photo current equivalent to that of the dark current, which can be written as \cite{Gopal2014}:
 \begin{equation}
NEP =\frac{ I_{0}\left[e^{qV_A/kT}-1 \right]}{\gamma *\sqrt{B}}
\label{NEP}
\end{equation}
where the term in the numerator is the dark current, $\gamma$ is the quantum efficiency  of the PIN diode, and  $I_0 =\frac{q A n_i^2}{N_A}\left[\frac{kT \mu_n}{q \tau_n}\right]^{1/2}\tanh(\frac{t}{L_n})$ is the saturation current, applied in the equation as its root mean square value. Here $N_A $ is the electrically active acceptor/hole concentration on the lightly doped p-side of the junction, $n_i$ is the intrinsic carrier concentration, $A$ is the junction area, $\tau_n$ is the minority carrier lifetime, $\mu_n$ is the minority carrier mobility, $L_n$ is the minority carrier diffusion length, $t$ is the thickness of the p-type base of the diode, and $V_{bi}$ is the bias voltage across the diode. $T$, $q$ and $k$ are, respectively, the temperature, the charge of the electron and Boltzmann's constant.

\begin{figure}[htb!]
 \includegraphics[width=0.5\textwidth]{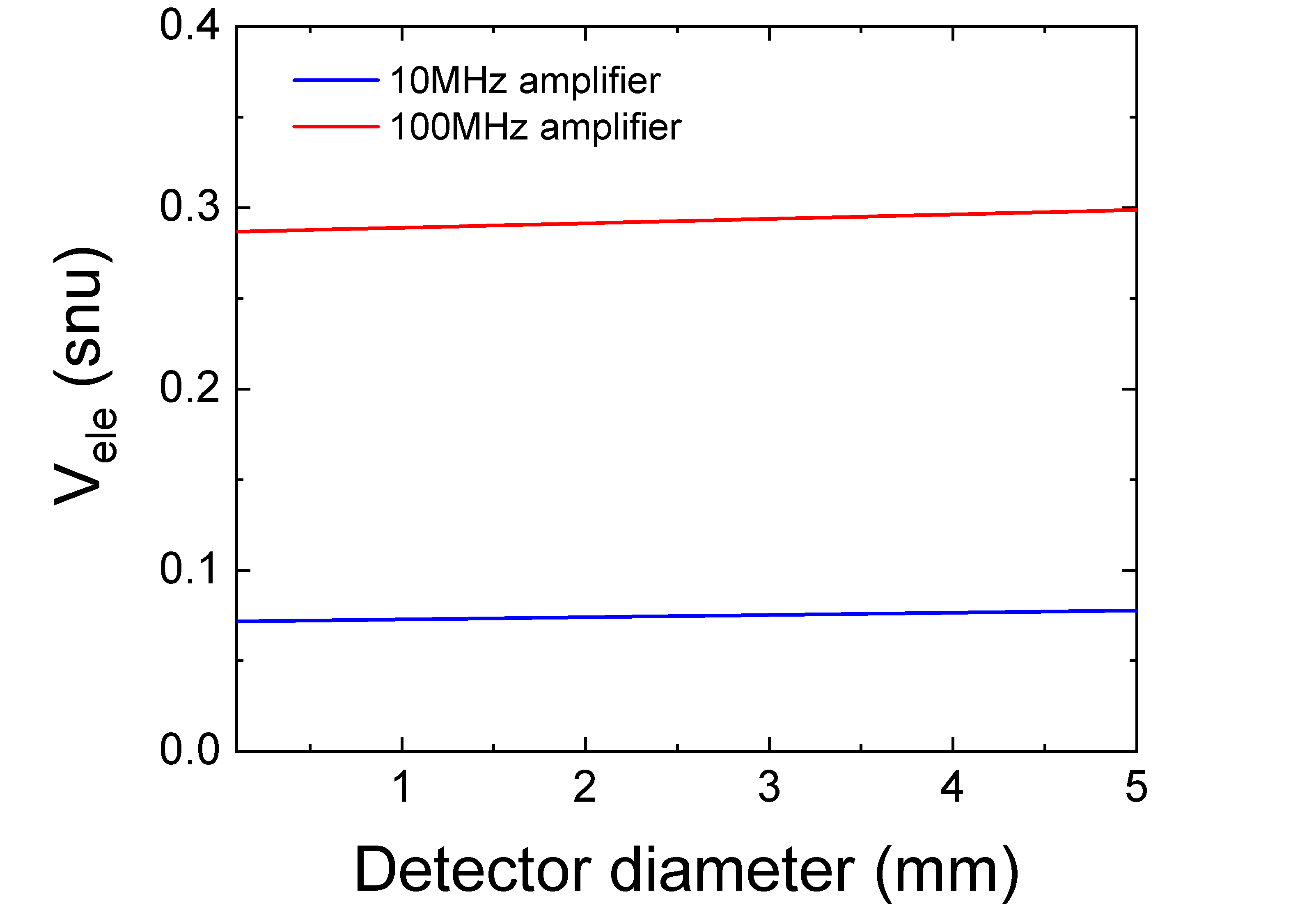}
\caption{ Variation of electronic noise variance  with respect to the detector diameter. The blue and red lines represent the cases of  10 MHz  and 100 MHz bandwidth for the amplifiers. The following values are used in Eq. \eqref{BW}:  $NEP_{amp}=5*10^{-12}\ W/ \sqrt{Hz} \text{ and } 10^{-11}\  W/ \sqrt{Hz}$; $\tau=0.3*B^{-1}$; $I_{lo}=13\ \mu W$.}
\label{Vele}     
\end{figure}

In figure \ref{BW}  the red line shows the variation of the bandwidth of the photo-diode with respect to the diameter. The bandwidth primarily reduces due to the increase in junction capacitance of the photo-diode with detection area.  The blue line shows the variation of $NEP$ of the photo-diode. It is evident from the graph that the noise from the photo-diode increases with respect to the diameter.  By inserting  Eq.\eqref{BW_Area} and Eq.\eqref{NEP} in Eq.\eqref{BW}, we can summarise that electronic noise variance increases with the detection area of the photo-diodes as shown in figure \ref{Vele}.

Since most of the CV-QKD systems are demonstrated at around a few MHz  clock rate,  we consider 10 MHz bandwidth amplifier with  $NEP=5*10^{-12}W/\sqrt{Hz}$. Even though the electronic noise increases and  overall bandwidth decreases, it is still manageable to achieve MHz clock rates with a 3 mm diameter detector, see figure \ref{BW}, under the assumption that the clock rate is 1/3  of the homodyne bandwidth. We have also provided an example of 100 MHz with a $NEP=10^{-11}W/\sqrt{Hz}$ amplifier. As expected, the electronic noise variance is higher compare to 10 MHz, due to higher amplifier $NEP$. Note that  the maximum clock rate, 33 MHz at 1/3 of the total bandwidth, cannot be achieved with a detector diameter larger than 1 mm.

\section{Performance of large area photodiode homodyne detector}
We have seen in the previous section that increasing the detection area of the  photo-diode decreases the detection bandwidth and increases the electronic noise variance.  In this section we will analyse the advantages of using a large area detector in a free-space CV-QKD link. Transmission of signal through a free-space channel creates  signal loss due to beam divergences. One way to decrease the signal divergence is to use a large aperture telescope at the transmitter as per the following relation with divergence angle $\Theta = 1.22 \lambda/D  $, where  $\lambda$ is the wavelength of the signal and $D$ is the diameter of the transmitter telescope\cite{Erik2018}. For transmission over short distances, the beam diameter can be maintained below a typical receiver telescope diameter (say, 40cm diameter) by using a suitably large transmitter aperture. For example, over a 10 km free-space link, an 8cm transmitter aperture  produces a 24cm beam diameter at the receiver, resulting in negligible diffraction losses. In the case of a satellite-to-ground link, as the transmitter aperture is limited by the size of the satellite, and in any case, impractically large transmitter and receiver apertures would be required due to the very large distance (hundreds of kilometres even for low Earth orbits).

In either case, in addition to any losses due to diffraction, atmospheric turbulence can cause further losses due to spreading and wander of the beam caused by spatially and temporally random variations in the refractive index of the air. In order to compare the  large area detector with fibre-coupled detectors, we have modelled the propagation of the signal beam through the atmosphere for the case of a satellite-to-ground link, and calculated the coupling losses for both types of detector. Atmosphere  turbulence is described statistically. The stochastic behaviour of refractive index is determined by the structure function $D_n(r) =  C_n^2 r^{2/3}$, which describes the expectation value of the difference in refractive index between two points a distance $r$ apart, where the term $C_n^2$ referred as refractive index structure parameter \cite{Andrews05}. $C_n^2$ is thus a measure of the strength of the turbulence, and is typically in the range $10^{-16}$ to $10^{-12}$ m$^{-2/3}$ at ground level. The strength of the turbulence decreases significantly with altitude, and in the present calculations we have used the Hufnagel-Vally model for the variation of $C_n^2$ with altitude \cite{Andrews05}.

\begin{figure}[htb!]
 \includegraphics[width=0.5\textwidth]{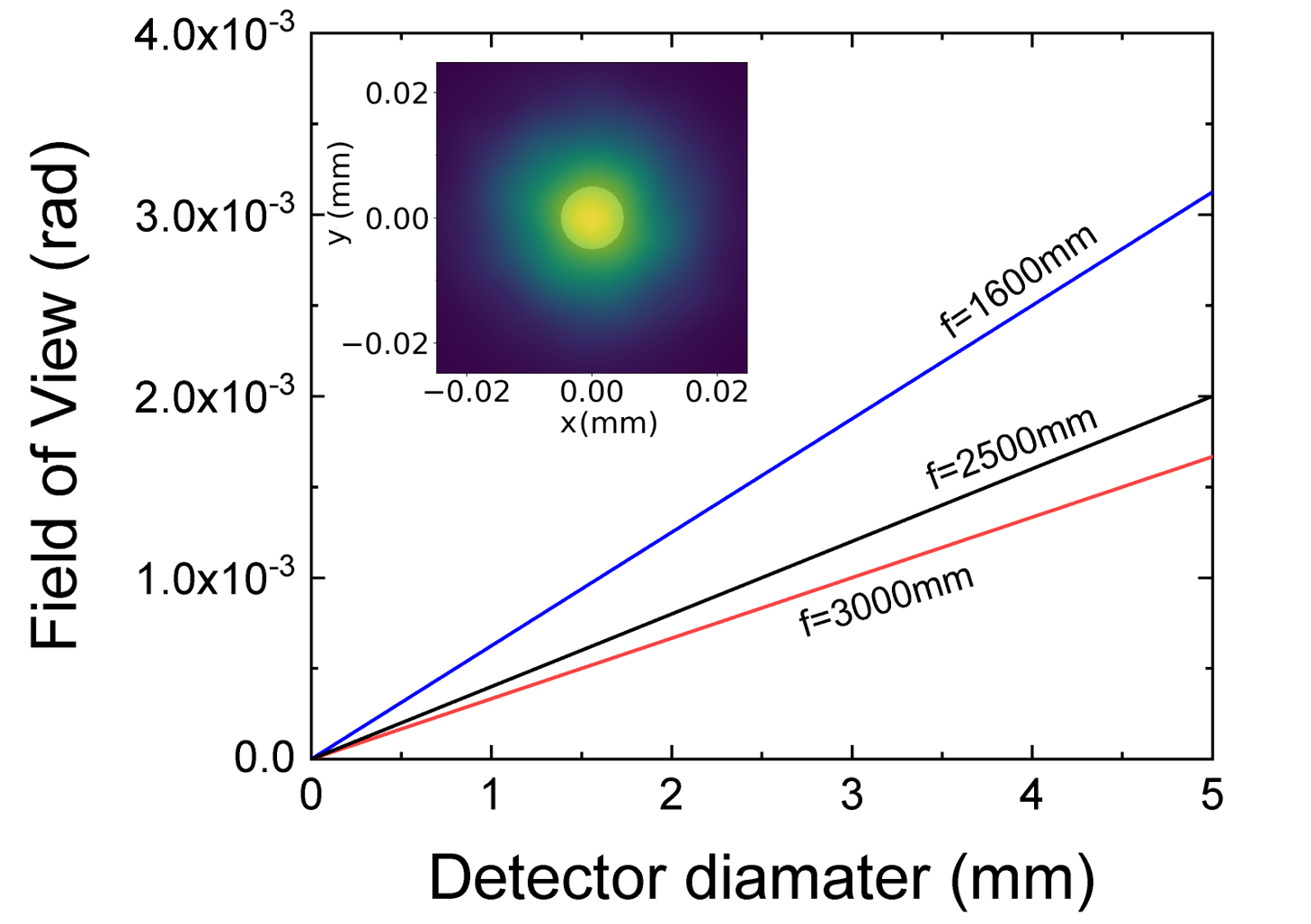}
\caption{Field-of-view  vs detector diameter. Field-of-view is estimated for different  focal length of the telescope.   The figure in the inset shows received signal intensity profile under moderate turbulence on a 10um core diameter of the single mode fibre (indicated by the paler circle at the centre of the figure inset). A large area detector easily collects all the light - the square region shows only the central 50 $\mu$m of the 1 mm detector.}
\label{FOV}     
\end{figure}

The case of a satellite in a 700 km low Earth orbit at an elevation of 90 degrees was considered. The 1550 nm signal beam was launched with a $1/e^2$ diameter of 7.1 cm, and a divergence of 14 microradians. Ground level turbulence of $10^{-13}$ $m^{-2/3}$ and a high altitude wind speed of 21 m s$^{-1}$ were assumed as inputs to the Hufnagel-Valley turbulence profile. The atmosphere was divided into 14 slices, each of which was modelled by propagation over the thickness of the slice and a random phase screen calculated based on Von Karmen statistics using the AOtools Python package. Fourier optics techniques were used to propagate the electric field to the plane of the detector/fibre facet, through the receiver, which had an aperture of 35.5 cm and an effective focal length of 1600 mm. A 10 $\mu$m diameter single-mode fibre and a square detector of side 1 mm were considered.

The inset of figure\ref{FOV} shows the  focal plane of the  receiver telescope with fibre  and  large area detectors under a weak turbulence conditions. It is visually evident that  the large area detector can collect more light and thereby reduces the signal loss.  One important thing to  note here is that, since the LO and signal  undergo identical turbulence effects,  the spatial  profile  of the signal is identical with that of the LO. As a result the signal is mode matched and coherent with respect to the LO - a requirement for efficient homodyne detection.

\begin{figure}[htb!]
 \includegraphics[width=0.5\textwidth]{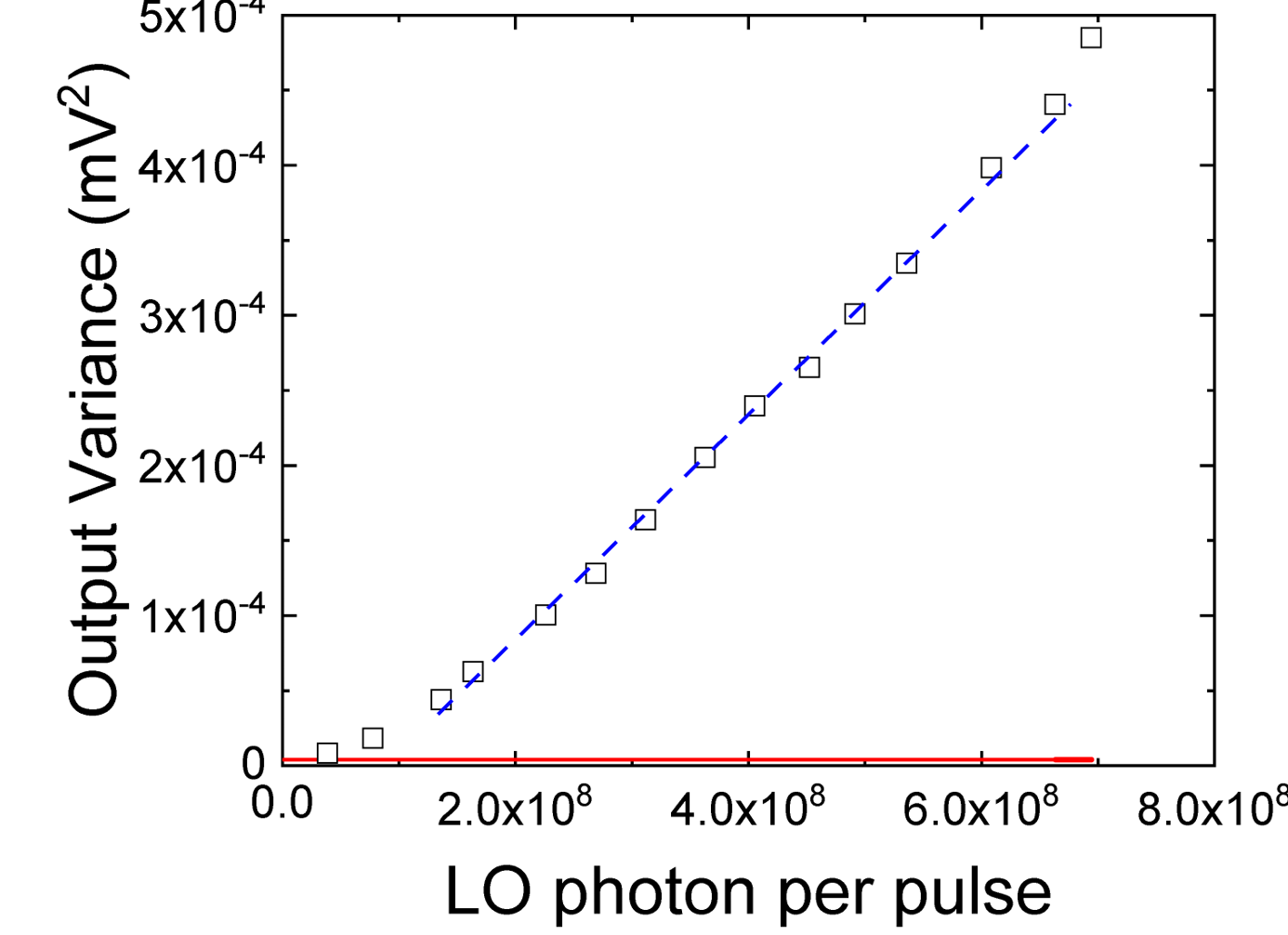}
\caption{Output variance of the homodyne detector at various local oscillator powers. The Blue dashed lines are added to indicate the linear region  for the detector. The red solid line is the electronic noise variance measured without the LO.}
\label{SN}     
\end{figure}

The average losses due to turbulence alone were 13.5 dB in the case of the fibre, while the large area detector easily collected all the light passing through the receiver aperture, with 0 dB of turbulence-induced losses - in that case only the diffraction losses of 31.8 dB were suffered. (Atmospheric absorption and scattering was not considered here.) We have also considered the field-of-view of the receiver  in terms of  detector area and telescopes focal length,$f$, given by $2\tan^{-1}(\sqrt{A/\pi}/f)$. The fibre facet has a field-of-view in this case of 6.25 microradians, while the 1mm detector has a field-of-view of 625 microradians for the present focal length. In practice, one could significantly increase the focal length of the receiver and still collect all or most of the signal light passing through the aperture, whilst reducing the field-of-view to minimise background noise. This gives substantially more design freedom. And also, larger field-of-view reduces the resolution requirement for the mount for tracking the transmitter.

In order to evaluate the shot-noise sensitivity of the large area detector based  homodyne detector, we have performed the following test. A 30ns wide laser pulse of wavelength 1550nm at 1MHz rate is used as the LO pulse  for the homodyne detector with 1mm diameter photo-diodes.  The photo-currents from both photo-diodes are made equal by balancing the homodyne detector. The output  variance is measured, without  signal, at various LO intensities.  Electronic noise variance of the detector is also measured without the LO. The measurement result is  shown in figure \ref{SN}.  The homodyne detector shows linear response with respect to the LO power. Electronic noise variance of 0.01$\%$ of the vacuum noise variance is observed at  $5\times10^8$ photon per pulse which can be achievable by a 1W laser over a 40dB atmospheric channel.

\section{Conclusion} 
We have shown that using large area photo-diodes  increases the signal collection efficiency  under atmospheric turbulence compared to a fibre-based  signal receiver. Considering loss from beam divergences and atmospheric turbulence, a fibre based receiver experiences 45.3dB loss while 1mm detection shows 31.8dB loss from a 700km satellite to ground link at 90 degrees.  Our homodyne receiver evaluation shows linear shot-noise limited sensitivity at reduced bandwidth, however still adequate for performing CV-QKD at typical MHz clock rates.  Such a receiver will also reduce the need for precision telescopic mount controls for a static link where transmitter and receiver are stationary or station-keeping. A noticeable feature of the  large area detector based homodyne detector is that the it obviates the need for adaptive optical elements in the free-space link. Larger field-of-view is  another benefit of using large area photo-diodes, however, this increases the background noise photon flux. Since the LO works as a mode selector for the quantum signals, higher FOV may not have a significant effect on CV-QKD systems. However, the effect of comparatively lower LO power may reduce the noise photon rejection ability of the homodyne detector. We will analyse the effect of large FOV in  secure key rate in our future study.

\section*{Acknowledgement}  This work has been funded by the Innovate UK project 3QN. R.K and T.S acknowledge the support from EPSRC via the UK Quantum Communications Hub (Grant No. EP/T001011/1).

\bibliography{cvref}

\end{document}